\documentclass[12pt,preprint]{aastex}                         
\begin{document}                                              
                                                              
\def\Tcol{$T_{\rm col}$}                                      
\def\fcol{$f_{\rm col}$}                                      
\def\Teff{$T_{\rm eff}$}                                      
\def\NH{$N_{H}$}       
\def\Lbol{$L_{\rm bol}$}                                      
\def\Mdd{$M_{\rm dd}$}                                        
\def\fdg{\hbox{$.\!\!^\circ$}}                                
                                                              
\title{Estimating the Spin of Stellar-Mass Black Holes via    
Spectral Fitting of the X-ray Continuum}                             

\author{Rebecca Shafee\altaffilmark{1},                       
Jeffrey E. McClintock\altaffilmark{2},                        
Ramesh Narayan\altaffilmark{2},                               
Shane W. Davis\altaffilmark{3},                               
Li-Xin Li\altaffilmark{4},                                    
Ronald A. Remillard\altaffilmark{5}}                          
                                                              
\altaffiltext{1}{Harvard University, Department of Physics,   
17 Oxford  Street, Cambridge, MA 02138}
\altaffiltext{2}{Harvard-Smithsonian Center for Astrophysics, 60 Garden 
Street, Cambridge, MA 02138}                                                   

\altaffiltext{3}{Department of Physics, University of California, Santa        
Barbara, CA 93106}                                                             

\altaffiltext{4}{Max-Planck-Institut f\"ur Astrophysik,                      
Karl-Schwarzschild-Str. 1, Postfach 1317, 85741 Garching, Germany}             

\altaffiltext{5}{Center for Space Research, Massachusetts Institute of 
Technology, Cambridge, MA 02139}                                               

\begin{abstract}                                                               

We fit X-ray spectral data in the thermal dominant or high soft state
of two dynamically confirmed black holes, GRO~J1655-40 and 4U~1543-47,
and estimate the dimensionless spin parameters $a_*\equiv a/M$ of the
two holes.  For GRO~J1655-40, using a spectral hardening factor
computed for a non-LTE relativistic accretion disk, we estimate $a_*
\sim 0.75$ and $\sim 0.65-0.75$, respectively, from {\it ASCA} and
{\it RXTE} data.  For 4U~1543-47, we estimate $a_* \sim 0.75-0.85$
from {\it RXTE} data.  Thus, neither black hole has a spin approaching
the theoretical maximum $a_*=1$.

\end{abstract}                                                                 


\normalsize                                                                   
                                                                               
\section{INTRODUCTION}      
   
An astrophysical black hole is completely defined by two numbers that
specify its mass and spin. The masses of 20 accreting black holes
located in X-ray binary systems have been determined or constrained by
dynamical optical studies (McClintock \& Remillard 2005, hereafter
MR05; Casares et al. 2004; Orosz et al. 2004). However, no reliable
measurements have been reported so far of the dimensionless black hole
spin parameter $a_* \equiv a/M$, where $a = J/cR_g$, $J$ is the
angular momentum of the black hole, $M$ is its mass, and $R_g=GM/c^2$.

In this work, we estimate $a_*$ of two black hole binaries by fitting
their X-ray thermal continuum spectra using a fully relativistic model
of a thin accretion disk around a Kerr black hole (Li et
al. 2005). The model includes all relativistic effects such as
frame-dragging, Doppler boosting, gravitational red shift, and bending
of light by the gravity of the black hole. It also includes
self-irradiation of the disk, the effects of limb darkening, and a
spectral hardening factor \fcol~to relate the color temperature
\Tcol~and the effective temperature \Teff~ of the disk emission:
\fcol~=~\Tcol/\Teff ~(Shimura \& Takahara 1995, hereafter ST95;
Merloni, Fabian \& Ross 2000; Davis et al. 2005, hereafter D05).  

In order to estimate the black hole spin by fitting the broadband
X-ray spectrum, one must know the mass $M$ of the black hole, the
inclination $i$ of the accretion disk (which we assume is the same as
the inclination of the binary system, but see Maccarone 2002), and the
distance $D$ to the binary (Zhang, Cui \& Chen 1997).  Accordingly, we
have selected two black hole binaries, GRO~J1655-40 (hereafter J1655)
and 4U~1543-47 (hereafter U1543), for which all three of these
quantities have been well determined from optical observations: for
J1655, $M = 6.30\pm0.27 M_\odot$, $i=70\fdg2\pm1\fdg2$, $D=3.2\pm0.2$
kpc, and for U1543, $M = 9.4\pm1.0 M_\odot$, $i=20\fdg7\pm1\fdg5$,
$D=7.5\pm1.0$ kpc (Orosz et al. 2003, and private communication;
Hjellming \& Rupen 1995).  

\section{OBSERVATIONS AND DATA REDUCTION}

We consider only those X-ray data that were obtained in the thermal
dominant state, or TD state (formerly high soft state), for which $>
75$\% of the 2--20 keV flux is supplied by the accretion disk (MR05).
This state is consistent with a simple multicolor blackbody model
(Gierlinski \& Done 2004), making it amenable to theoretical modeling.
We carried out all data analysis and model fits using XSPEC {\it
version 12.2.0} and HEASOFT {\it version 5.2}.

The continuum X-ray spectrum of J1655 was observed by the {\it
Advanced Satellite for Cosmology and Astrophysics (ASCA)} on 1995
August 15 (Ueda et al. 1998; hereafter U98) and on 1997 between
February 25 and 28 (Yamaoka et al. 2001; hereafter Y01), with total
exposure times of $\approx$19 ks and $\approx$100 ks, respectively.
The source was very bright during both observations (2.3 Crab in 1995
and 1.2 Crab in 1997), and we therefore analyzed only the data from
the GIS2 and GIS3 detectors.  Starting with the unscreened {\it ASCA}
data files obtained from the HEASARC, we followed as closely as
possible the data reduction procedures and criteria mentioned in U98
and Y01.
As in Y01, we added a systematic error of 2\% to the
GIS spectra to account for calibration uncertainties.  As a check on
our reduction procedures, we fitted both the 1995 and 1997 spectra
using the disk blackbody models employed by U98 and Y01, and we
succeeded in closely reproducing all of their published results.
  
J1655 was also observed with the {\it Rossi X-ray Timing Explorer
(RXTE)} in 1997 (Sobczak et al. 1999), including an observation on
February 26 which was performed simultaneously with ASCA.  Our second
source, U1543, was observed with {\it RXTE} during its 2002 outburst
(Park et al. 2004; hereafter P04).  For both {\it RXTE} data sets, the
data reduction procedures are identical to those described by P04.  In
brief, we used the ``Standard 2 mode'' data from PCU-2 only. The event
files and spectra were screened and the background spectra and a
response files created.  Systematic errors of 1\% were added to all the
PCU-2 energy channels.  
Referring to Figure 2 in Remillard (2005), we selected the contiguous
group of 31 observations extending from MJD 50453.6 to MJD 50663.7 for
which the source J1655 was in the TD state.  Because the data span 210
days, we created and used several different response files.  For
U1543, from among the 49 observations considered by P04 (see their
Table 1), we selected the 34 observations (Obs. nos. 1-3, 5-19, 27-42)
for which the source was in the TD state (MR05).
                                                                               
\section{DATA ANALYSIS}                                                        

{\it RXTE long fits}: First we fitted the {\it RXTE } pulse-height
spectra of J1655 and U1543 in the 2.8--25.0 keV range.  These fits,
which we refer to as the ``long fits,'' were made using a spectral
model comprised of three principal components: {\it kerrbb}, which
models a relativistic accretion disk (Li et al. 2005), a standard
low-energy absorption component ({\it phabs}), and a simple power-law
component ({\it power}).  In addition, following the work of Y01, we
found it necessary to add three edge/line features in order to obtain
acceptable fits: (1) a smeared Fe edge ({\it smedge}) with edge energy
restricted to the interval 6.8--9.0 keV and width fixed at 7 keV; (2)
a sharp absorption edge with edge energy restricted to the interval
9--11 keV; and (3) a Gaussian absorption line that was used solely for
J1655.  The central energy of this line was restricted to the range
6.4--7.0 keV, and its width was fixed at 0.5 keV.  (The width was
determined by our analysis of the {\it ASCA} data, see below).  We
fixed the equivalent neutral hydrogen column density \NH~at $0.7
\times 10^{22} {\rm cm}^{-2}$ for J1655 (Y01) and $0.4 \times 10^{22}
{\rm cm}^{-2}$ for U1543 (P04).

{\it RXTE short fits}: Since we are interested in the disk component
of the spectrum, we analyzed the same data over restricted energy
ranges that are dominated by the thermal component, thereby generating
``short fits.''  For all these short fits, we used the three principal
componenets used in the long fits. For the J1655 data we also used the
gaussian line feature mentioned above.  We explored several different
upper energy limits for the {\it RXTE} spectra, finally choosing
2.8--7.5 keV for J1655 and 2.8--7.0 keV for U1543.  We determined the
upper limit by the highest energy that allowed the exclusion of the
{\it smedge} component required by the long fits (for which this
feature had a mean edge energy of 7.7 keV for J1655 and 7.0 keV for
U1543).  The 7.5 keV limit for J1655 (vs. 7.0 keV) was required to
ensure that the parameters of the Fe absorption line feature present
between 6.4 and 7.0 keV were well-determined.  Twenty-six out of 31 of
the J1655 short fits (including the 1997 Feb 26 spectrum featured in
Fig. 1) succeeded without a power-law component (i.e., $\chi_\nu^2 <
1$); furthermore, its inclusion did not improve these fits
significantly.  However, in the case of U1543 (which on average had a
power-law to total flux ratio of 0.15, compared to 0.06 for J1655),
the power-law component was essential in fitting all 34 of the
spectra. In order to be consistent, the power-law component was
included in all short fits of both sources with parameters fixed to
the values determined from the long fits.

{\it ASCA short fits}: The {\it ASCA} data for J1655 were analyzed
only over a short energy range, which was chosen to be 1.2--7.5 keV,
because the bandwidth of the GIS detectors is too limited to constrain
the power-law component.  For 1997 {\it ASCA} we modelled the spectra
using {\it kerrbb}, {\it phabs} and the gaussian absorption line mentioned
earlier. For the 1995 data we used only {\it kerrbb} and {\it phabs}.

In the analysis of the {\it RXTE} data, we found it necessary to
correct the fluxes downward as follows.  For U1543, the Crab flux
calculated using the P04 response file exceeded the flux predicted by
the standard Crab spectrum of Koyama et al. (1984) by a factor of
1.174.  For J1655, we used a current response matrix and found a much
smaller correction factor of 1.034.

In fitting the data using {\it kerrbb}, we fixed $M$, $i$ and $D$ to their
mean observed values (\S1). Also, we switched on limb-darkening
(lflag=1) and returning radiation effects (rflag=1).  We set the
torque at the inner boundary of the accretion disk to zero, fixed the
normalization to 1 (as appropriate when $M$, $i$ and $D$ are held
fixed), and allowed the mass accretion rate $\dot M$ to vary freely.
Of the two remaining parameters, viz., the spectral hardening factor
\fcol~and the black hole spin $a_*$, we held one or the other fixed
and fitted the other.

\section{RESULTS}                                                              

Figure 1 shows the results of analyzing the 1995 and 1997 {\it ASCA}
data on J1655 with {\it kerrbb}.  In this analysis, we kept \fcol~fixed at
selected values and fitted $a_*$ and $\dot M$.  We see that, for a
given value of \fcol, the data are able to determine $a_*$ accurately.
The $\chi_\nu^2$ values are acceptable, and there is  reasonable
agreement between the GIS2 and GIS3 results and between the 1995 and
1997 data.  There is agreement also with the simultaneous {\it RXTE}
observation done on 1997 Feb 26.  However, since the $\chi_\nu^2$ is
acceptable for all the values of \fcol, it is clear that the data by
themselves cannot constrain \fcol.  We thus need an independent
theoretical determination of \fcol~if we wish to estimate $a_*$.

ST95 were among the first to calculate model disk atmospheres for
black hole binaries, including full radiative transfer and
Comptonization.  From these models they estimated \fcol~as a function
of the bolometric disk luminosity $L_{\rm bol}$.  Roughly, they found
$f_{\rm col} \approx 1.7 + 0.2(\log\ell+1.25)$, where $\ell \equiv
L_{\rm bol}/L_{\rm Edd}$ is the Eddington-scaled disk luminosity and
$L_{\rm Edd} = 1.5 \times 10^{38} \; (M/M_{\odot})$.  Using the
observed luminosities of J1655 during the {\it ASCA} 1995 and 1997
observations, viz., $\log\ell \sim -0.85, ~-1.02$, we have estimated
the respective values of \fcol~ according to the ST95 model.  These
are shown by the dotted lines marked ST95 in Figure 1.  The last
column of Table 1 shows the corresponding estimates of $a_*$.

Recently, D05 computed more detailed disk atmosphere models which
improve upon the pure hydrogen atmospheres of ST95 by including metal
opacities.  Metals tend to reduce the spectral hardening.  Therefore,
the D05 values of \fcol~ are generally smaller than those of ST95.
For the analysis presented in this paper, we used D05's code to
calculate a grid of values of \fcol~as a function of $\ell$ and $a_*$.
The calculations were done for the specific inclinations of J1655 and
U1543, and for two values of the disk viscosity parameter, $\alpha=$
0.01, 0.1.  Figure~1 shows the range of \fcol~values predicted by the
D05 model for the 1995 and 1997 {\it ASCA} observations, and the
second to last column of Table 1 gives the corresponding estimates of
$a_*$.  While we present in this paper results for both the ST95 and
D05 models of spectral hardening, we view the latter as more reliable.

Figure 2 shows the short-fit results (\S 3) of all the 31 {\it RXTE}
observations of J1655 in the TD state.  Here we have assumed various
values of $a_*$, and computed for each observation the best-fit values
of \fcol~and $\ell$ (the latter is obtained from $\dot M$ and $a_*$).
We then compare the data-fitted values of \fcol~with the model
predictions of D05 and ST95.  The comparison with the D05 model
indicates that the spin of J1655 is likely to lie in the range $a_*
\sim 0.65-0.75$; this result is entered in Table 1.  It is interesting
to note that the D05 model gives nearly identical results for
$\alpha=0.01$ and 0.1 so long as $\log\ell \lesssim -1$, but shows
noticeable variations with $\alpha$ when the disk is more luminous.
Since the true $\alpha$ of the disk is not known (it is likely to be
in the range 0.01 to 0.2), and moreover since disks tend to be
radiation pressure dominated at higher luminosities thereby
introducing additional uncertainties, we give greater weight to the
observations for which $\log\ell<-1$.  Thus, we ignore the 1995 {\it
ASCA} data ($\log\ell\sim-0.85$), and give more weight to the 1997
data ($\log\ell\sim-1.02$). We also favor the GIS2 results because the
GIS3 spectra have a prominent feature between 1 and 2 keV that cannot
be eliminated either by using the XSPEC ``gain'' command and/or by
adding an ad hoc absorption edge to the fit.

Figure 3 shows a similar analysis for U1543. Here we found that 7 of
the 34 observations in the TD state (MJD 52444.515-52449.11) gave poor
fits for the short range due to an edge-like feature between 4 and
5 kev which is not present in other observations. The long fits
for these data still gave reasonable fits with $\chi_\nu^{2} <
2$. Comparing the fitted values of \fcol~ with the D05 model, we
estimate $a_*\sim 0.75-0.85$. Once again we focus on the 
lower-luminosity data with $\log\ell < -1$.

The disk parameters obtained from the short and long fits agree very
well in all cases, with typical differences in $\log\ell$ and $f_{\rm
col}$ of order $0.003$ and $0.005$ respectively; the uncertainties in
the fitted values are considerably larger ($\sim 0.01$ and $\sim
.02$).  Also to check whether the different energy ranges of the {\it
ASCA} and {\it RXTE} data is an issue, we reanalyzed the {\it ASCA}
data using only the restricted energy range 2.8--7.5 keV and found the
results agree.  We also explored the effect of varying the values of
$M$, $D$ and $i$ over one standard deviation in either direction (\S
1).  We find that the derived $a_*$ values lie within the ranges given
in Table 1. We note that system parameters need to be measured with
high accuracy for this method to succeed.  For example, we attempted
to analyze the BH candidate XTE J1550-564 using the methods described
here, but since the distance to this source is quite uncertain, $D =
5.9^{+1.7}_{-3.1}$ (Orosz et al. 2002), we were unable to obtain any
useful constraint on $a_*$.

\section{SUMMARY AND CONCLUSIONS}                                              
                                                                               
The method of estimating spin that we have employed was pioneered by
Zhang et al. (1997; see also Gierlinski et al. 2001).  However, only
recently have the necessary data analysis tools ({\it kerrbb}, Li et
al. 2005) and disk atmosphere models (D05) been developed to the point
where the method may be applied with some confidence.
                                                                               
Effectively, in this technique, one determines the radius $R_{\rm in}$
of the inner edge of the accretion disk and assumes that this radius
corresponds to the innermost stable circular orbit ($R_{\rm ISCO}$).
Since $R_{\rm ISCO}/R_g$ is a monotonic function of $a_*$, a
measurement of $R_{\rm in}$ and $M$ directly gives $a_*$.  Provided
that (i) $i$ and $D$ are known to sufficient accuracy, (ii) the X-ray
flux and spectral temperature are measured from well calibrated X-ray
data in the TD state, and (iii) the disk radiates as a blackbody, it
is clear that $R_{\rm in}$ can be estimated.  However, the disk
emission is not a true blackbody but a modified blackbody with a
spectral hardening factor \fcol.  Therefore, the observations only
give the quantity $R_{\rm in}/f_{\rm col}^{2}$, and we need an
independent estimate of \fcol~in order to estimate $a_*$.  
We have tailored the state-of-the-art disk
atmosphere model of D05 to obtain estimates of \fcol~for this work.
                                                                               
Our results in brief are as follows.  By fitting {\it ASCA} and {\it
RXTE} spectral data on the black hole X-ray binary J1655, we estimate
the dimensionless spin parameter of the black hole to be $a_* \sim
0.65-0.75$ (Table 1).  In the case of U1543 we estimate $a_* \sim
0.75-0.85$, though this is based on only {\it RXTE} data.  To obtain
these estimates, we have focused on observations for which the disk
luminosity was relatively low ($\log\ell < -1$), and we have used the
D05 model for \fcol~(though for completeness we give results also for
the ST95 model in Table 1 and in the Figures).
                                                                               
Based on these results, we consider it unlikely that either J1655 or
U1543 has a spin close to the theoretical maximum for a rotating black
hole, $a_*=1$.  Even $a_*=0.85$, which is the largest value we find,
corresponds to a quite moderate spin, as one can see by considering
the binding energy per unit mass of a particle in the last stable
circular orbit: For $a_*$ in the range 0 to 1, this quantity varies
from 5.7\% to 42.3\%, whereas it is only 13.6\% for $a_*=0.85$.
Moreover, most systematic effects that one might consider only push
our estimates of $a_*$ down.  If $\alpha$ is larger than 0.1, as
suggested by some studies of white dwarf disks, it would cause our
estimates of $a_*$ to decrease (see Figs. 2, 3).  Similarly, if we
allow a non-zero torque at the inner edge of the disk and/or allow the
disk to radiate inside $R_{\rm ISCO}$ (Krolik 1999), $a_*$ would
decrease still further.
                                                                               
What spin might we expect a black hole to accrue due to disk accretion
alone?  If the hole accretes long enough to achieve spin equilibrium,
then the limiting $a_*$ is likely to be in the range $\sim 0.9-0.998$
(Gammie, Shapiro \& McKinney 2004, and references therein).  However,
X-ray binaries rarely live long enough for such equilibrium to be
established.  U1543, for instance, contains a $\approx2.5 M_{\odot}$
main-sequence secondary (Orosz et al. 1998), and thus the age of the
system is $\lesssim 1$~Gyr.  Based on the X-ray fluences from the
1971, 1983, 1992 and 2002 outbursts, the average mass accretion rate
is $\sim 1 \times 10^{-9} M_{\odot}$ yr$^{-1}$ for $D = 7.5$~kpc
(Chen, Shrader, \& Livio 1997; P04).  At this rate, the black hole
will accrete at most 1~$M_{\odot}$ during the lifetime of the system
and thus, assuming that its natal spin is zero, the spin today should
be $a_* \lesssim 0.35$, considerably less than our estimate of $a_*
\sim 0.75-0.85$.  This suggests that our measurements are sensitive to
the natal spins of these black holes.  It is then interesting that
neither of the two holes has a spin close to either 0 or 1.
                                                                               
\acknowledgments                                                               
                                                                               
We thank Ken Ebisawa for helpful discussions on {\it ASCA} data
analysis, Jean Swank and Keith Jahoda for information on the effective
area of the {\it RXTE} PCA, Keith Arnaud for help in implementing
{\it kerrbb} in XSPEC, and Saeqa Vrtilek and the referee for useful
comments.  This research was supported in part by NASA grant
NNG~05GB31G and NSF grant AST 0307433, and has made use of data
obtained from the High Energy Astrophysics Science Archive Research
Center (HEASARC), provided by NASA's Goddard Space Flight Center.

\clearpage                                                                     
                                                                               
\begin{figure}                      
  \figurenum{1}                                                                

\plotone{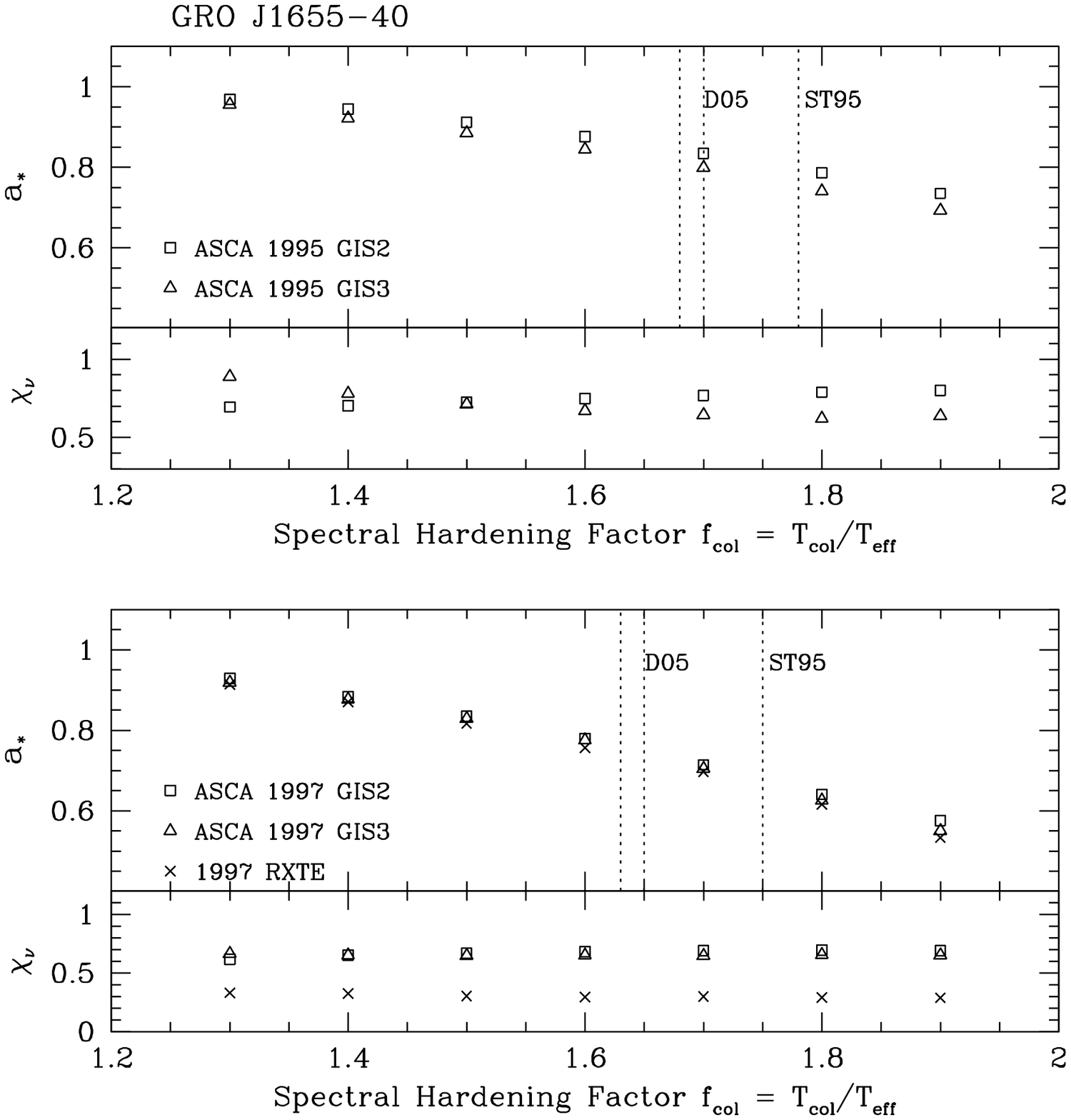}                                                               

\caption{[Top] GRO~J1655-40 {\it ASCA} 1995 spectra fitted with
different fixed values of the spectral hardening factor \fcol. The
upper panel shows the values of $a_*$ obtained for each value of
\fcol, and the lower panel shows the corresponding reduced
$\chi_\nu^{2}$ (84  dof). The vertical dotted lines indicate the
likely values of \fcol~ according to the D05 and ST95 models.
[Bottom] Corresponding results for {\it ASCA} 1997 and {\it
RXTE} 1997 Feb 26 spectra. Errorbars in $a_*$ are very small on
average ($\sim.005$) and are not shown for clarity.  The {\it ASCA}
fits are for  81 dof and the {\it RXTE} fits are for  8 dof.}
\end{figure}

\clearpage                                                                     
                                                                               
\begin{figure}                                                                 

\figurenum{2}                                                                  

\plotone{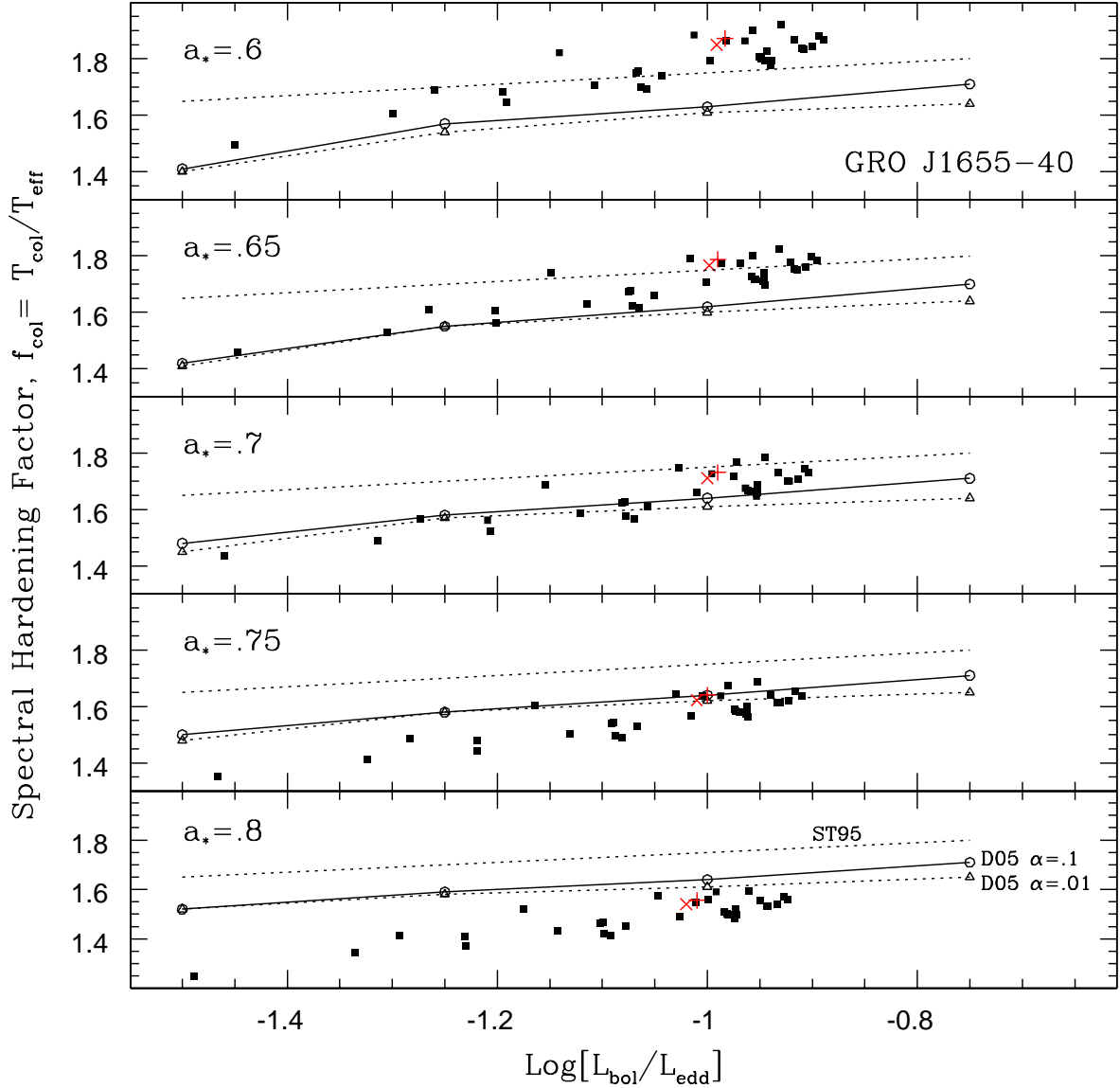}                                                               

\caption{J1655 {\it RXTE} spectra fitted for  five different
fixed values of $a_*$. The solid squares show the fitted values
of the spectral hardening factor \fcol~and the dimensionless
luminosity $\ell$ 
obtained with short fits.  The {\it ASCA} 1997
GIS2 point is shown as a red plus and the GIS3 point as a red
cross. The lines show the calculated values of \fcol~from the D05
model for $\alpha=0.01$ and 0.1 and from the ST95 model . The solid
line is for $\alpha=0.1$, the value we emphasize.  The D05 model
constrains $a_*$ of J1655 to lie in the range $0.65-0.75$.  Errorbars
in \fcol~ (typically $\sim0.02$) are not shown.  The fits are for 8
dof with $\chi_\nu^2$ typically $\sim0.3$.}

\end{figure}

\clearpage                                                                     
                                                                              
\begin{figure}                                                                 

\figurenum{3}                                                                  

\plotone{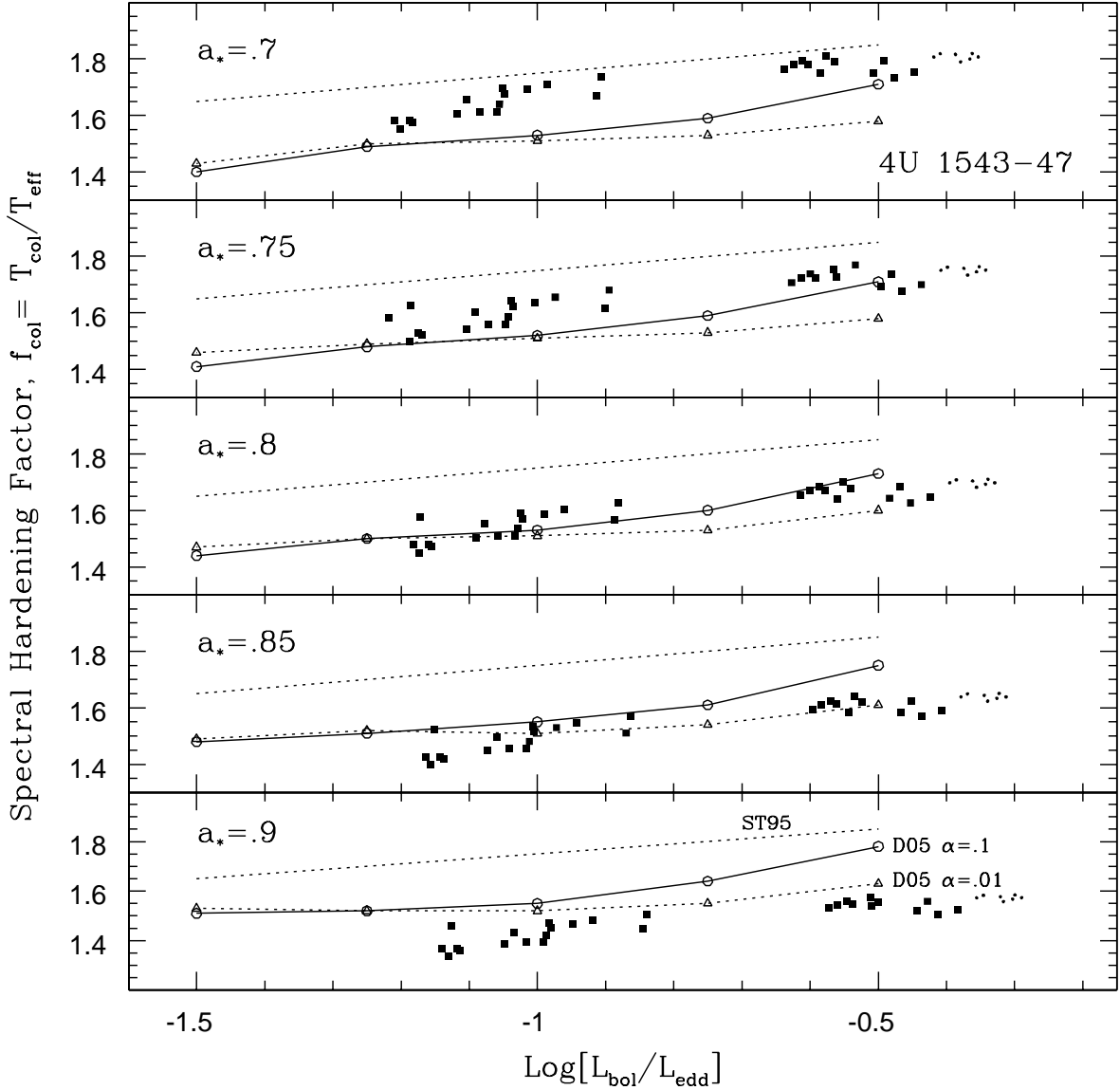}                                                               

\caption{ U1543 {\it RXTE} spectra fitted for five different fixed
values of $a_*$.  The data points and lines have the same meanings as
in Fig. 2. The 7 dots on the extreme right of each panel correspond to
spectra that contained an absorption edge like structure between 4 and
5 keV and gave fits with $\chi_\nu^2 >2$. The D05 model constrains
$a_*$ of U1543 to lie in the range $0.75-0.85$.  The fits are for 7
dof with $\chi_\nu^2$ typically $\sim0.3$.}

\end{figure}
                                                                               
\clearpage                                                                     
                                                                               
\begin{center}                                                                 

\begin{tabular}{cccccc}                                                        

\multicolumn{6}{c}{TABLE 1} \\ & \\ \multicolumn{6}{c}{Black Hole Spin         

Estimates Using The Mean Observed Values of $M$, $D$, and $i$} \\              

\hline \hline                                                                  
\multicolumn{1}{c}{Candidate}&\multicolumn{1}{c}{Observation                   
Date}&\multicolumn{1}{c}{Satellite}& 
\multicolumn{1}{c}{Detector}&\multicolumn{1}{c}{$a_*$(D05)}&
\multicolumn{1}{c}{$a_*$ (ST95) } \\ \hline GRO J1655-40 &1995 August    
15 &ASCA &GIS2 & $\sim0.85$ & $\sim0.8$ \\ GRO J1655-40&1995 August 15 
&ASCA &GIS3 & $\sim0.80$ & $\sim0.75$ \\ GRO J1655-40 &1997 February 
25-28 &ASCA &GIS2 &   $\sim {\bf 0.75}$   & $\sim0.70$ \\ GRO J1655-40 &1997 
February 25-28 &ASCA &GIS3 &  $\sim {\bf 0.75} $  & $\sim0.7$ \\ GRO J1655-40 
&1997 February 26 &RXTE &PCA &  $\sim {\bf 0.75 }$  & $\sim0.65$ \\ GRO
J1655-40                 
&$1997($several$)$ &RXTE &PCA & {\bf 0.65--0.75} & 0.55--0.65 \\ 4U 1543-47
&$2002($several$)$ &RXTE &PCA & {\bf 0.75--0.85} & 0.55--0.65 \\ \hline  
\end{tabular}   

\end{center}  
\end{document}